\documentclass[prl,10pt,tightenlines,amsmath,amssymb,amsfonts,twocolumn,floatfix,%
superscriptaddress,showpacs
]{revtex4}

\usepackage{amsthm}
\usepackage{graphicx}
\usepackage{color}
\usepackage{bm}
\usepackage{times}

\newcommand\arrow\vec

\usepackage[colorlinks=false, linkcolor=blue, citecolor=magenta, bookmarks=false]{hyperref}

\def\id{I}

\def\1{\mat{\id}}

\def\mat#1{\bm{\mathrm{#1}}}

\renewcommand{\vec}[1]{\bm{\mathrm{#1}}}

\def\controlled#1{\mathrm{C}_{#1}}
\def\CZ{\controlled Z}

\begin{document}

\title{Arbitrarily Large Continuous-Variable Cluster States from a Single Quantum Nondemolition Gate}

\author{Nicolas C. Menicucci}
\affiliation{Perimeter Institute for Theoretical Physics, Waterloo, Ontario N2L 2Y5, Canada}

\author{Xian Ma}
\affiliation{Department of Physics, The University of Queensland, St Lucia, Queensland 4072, Australia}

\author{Timothy C. Ralph}
\affiliation{Department of Physics, The University of Queensland, St Lucia, Queensland 4072, Australia}
\affiliation{Centre for Quantum Computer Technology, The University of Queensland, St Lucia, Queensland 4072, Australia}

\date{February 22, 2010}

\begin{abstract}
We present a compact experimental design for producing an arbitrarily large optical continuous-variable cluster state using just one single-mode vacuum squeezer and one quantum nondemolition gate.  Generating the cluster state and computing with it happen simultaneously: more entangled modes become available as previous modes are measured, thereby making finite the requirements for coherence and stability even as the computation length increases indefinitely.
\end{abstract}


\maketitle


\emph{Introduction.}---%
Continuous-variable one-way quantum computation (QC)~\cite{Menicucci2006} combines quantum information processing using continuous variables~(CV's)~\cite{Braunstein2005a} with the experimental simplicity of one-way quantum computation~\cite{Raussendorf2001} using cluster states.  A cluster state, whether CV~\cite{Zhang2006} or qubit-based~\cite{Briegel2001}, has no computational power on its own.  For it to be useful, one must make local projective measurements on it.  The measurement sequence depends on the algorithm to be implemented and on the outcomes of previous measurements.  For optical CV cluster states, homodyne detection alone is sufficient to implement all Gaussian operations, while the additional availability of photon counting (or another measurement in a non-Gaussian basis) allows the cluster state to be used for universal QC.  The graph for the cluster state~\footnote{While usage varies in the literature, we use the convention~\cite{Nielsen2006,Menicucci2006,Menicucci2008,Flammia2009} that the term `cluster state' is entirely synonymous with `graph state.'} must also be sufficiently connected and scale appropriately with the number of qumodes and length of the computation.  In an optical setting, CV one-way QC has an advantage over qubit methods~\cite{Nielsen2004,Browne2005,Duan2005} (which are all nondeterministic) because creating optical CV cluster states is an entirely deterministic process.  Reference~\onlinecite{Gu2009} has further details on CV one-way QC.

The optical method of construction proposed in the original paper~\cite{Menicucci2006}, referred to here as the \emph{canonical method}, requires single-mode squeezers and controlled-$Z$ ($\CZ$)~gates, which are a special type of quantum nondemolition (QND) interaction~\cite{Bachor2004}.  The $\CZ$~gate can be implemented using beamsplitters and inline squeezing (i.e.,\ squeezing of a state other than the vacuum)~\cite{Braunstein2005,Yurke1985}, which is experimentally challenging but achievable using current technology~\cite{Yoshikawa2008}.  With many such gates required for a typical cluster-state computation~\cite{Gu2009}, more efficient methods for generating CV cluster states are desired.

One such proposal eliminates the need for $\CZ$~gates altogether by replacing them with a suitable beamsplitter network~\cite{vanLoock2007}.  Another method uses frequency encoding and produces large CV cluster states using just a single optical parametric oscillator (OPO)~\cite{Menicucci2008,Flammia2009}.  While both proposals represent significant advances over the canonical method, they both suffer from drawbacks that affect their scalability.  The beamsplitter-only approach~\cite{vanLoock2007} suffers from a need for extra squeezers and a continually larger (but still interferometrically stable!)\ beamsplitter network as the size of the cluster state grows.  Still, this method is amenable to proof-of-principle experiments with existing technology~\cite{Su2007,Yonezawa2008,Yukawa2008,Ukai2010}.  The single-OPO approach~\cite{Menicucci2008,Flammia2009} suffers from a need to pack more and more frequencies within the phasematching bandwidth of the OPO and the need for frequency-sensitive measurements.  Still, this method does not suffer from an increase in size of the physical experiment in any way comparable to that hindering the beamsplitter-only method.  Thus, while the initial implementation is more complex, once the technology is established~\cite{Pooser2005,Pysher2009,Pysher2009a,Midgley2010}, it is in principle much easier to scale up to cluster states larger by several orders of magnitude~\cite{Flammia2009}.

Despite these advances and their relative merits, both optical methods still assume that the entire state is prepared ahead of time and its coherence maintained during the adaptive measurement process.  It would be much more useful to have a cluster state that is extended as needed---simultaneously with measurements implementing an algorithm on it---in a manner analogous to repeatedly laying down additional track in front of a moving train car~\footnote{Occasionally such a method of one-way quantum computing is referred to as a ``Wallace and Gromit'' approach, alluding to the 1993 animated film, \emph{The Wrong Trousers} (one of several such films starring the two characters, Wallace and Gromit), in which Gromit, while traveling at high speed aboard a model train, reaches the end of the existing track and is forced to repeatedly place additional track pieces in front of him in order to avoid derailing and thwarting his pursuit of an outlaw penguin.}.  Such a method eliminates the need for long-time coherence of a large cluster state because only a small piece of the state exists at any given time.  (Reference~\onlinecite{Devitt2009} details a proposal using this idea for optical qubits.)

Implementing such a method for CV cluster states using the beamsplitter-only method is difficult because, unlike the flexibility afforded by $\CZ$~gates (which commute and can therefore be performed in any order~\cite{Gu2009}), all individual parameters of the interferometer are chosen as a whole to make the entire cluster state, and in general, extending it---even by one node---will require an entirely different interferometer.  The single-OPO method suffers similar difficulties since scalability of the method is ensured by packing all of the necessary interactions into the OPO design and generating a large cluster state all at once within a single output beam.  In this method, as well, a simple way of extending it is not known.

The canonical method was superseded soon after its inception because it is inefficient in its use of squeezing resources~\cite{vanLoock2007,Gu2009} and because $\CZ$~gates are experimentally challenging.  This is an important consideration if one $\CZ$~gate is required for every link between nodes, as originally proposed~\cite{Menicucci2006}.  But we show in this paper that with a temporal-mode encoding these considerations are much less important because the optical setup is reduced to just \emph{one single-mode squeezer} and \emph{one $\CZ$~gate}, regardless of size of the cluster state.  This means that, for an $N$-node cluster state, we don't need $N$~high-strength single-mode squeezers anymore; we only need one.  We also don't need $O(N^2)$ low-noise $\CZ$~gates; we only need one.  In what follows, we show how this works, starting with a linear CV cluster state (a quantum wire) of arbitrary length and then generalizing to a square-lattice CV cluster state of arbitrary breadth and depth.  The latter can be used for universal one-way QC~\cite{Gu2009}.


\begin{figure}
\begin{center}
\includegraphics[width=\columnwidth]{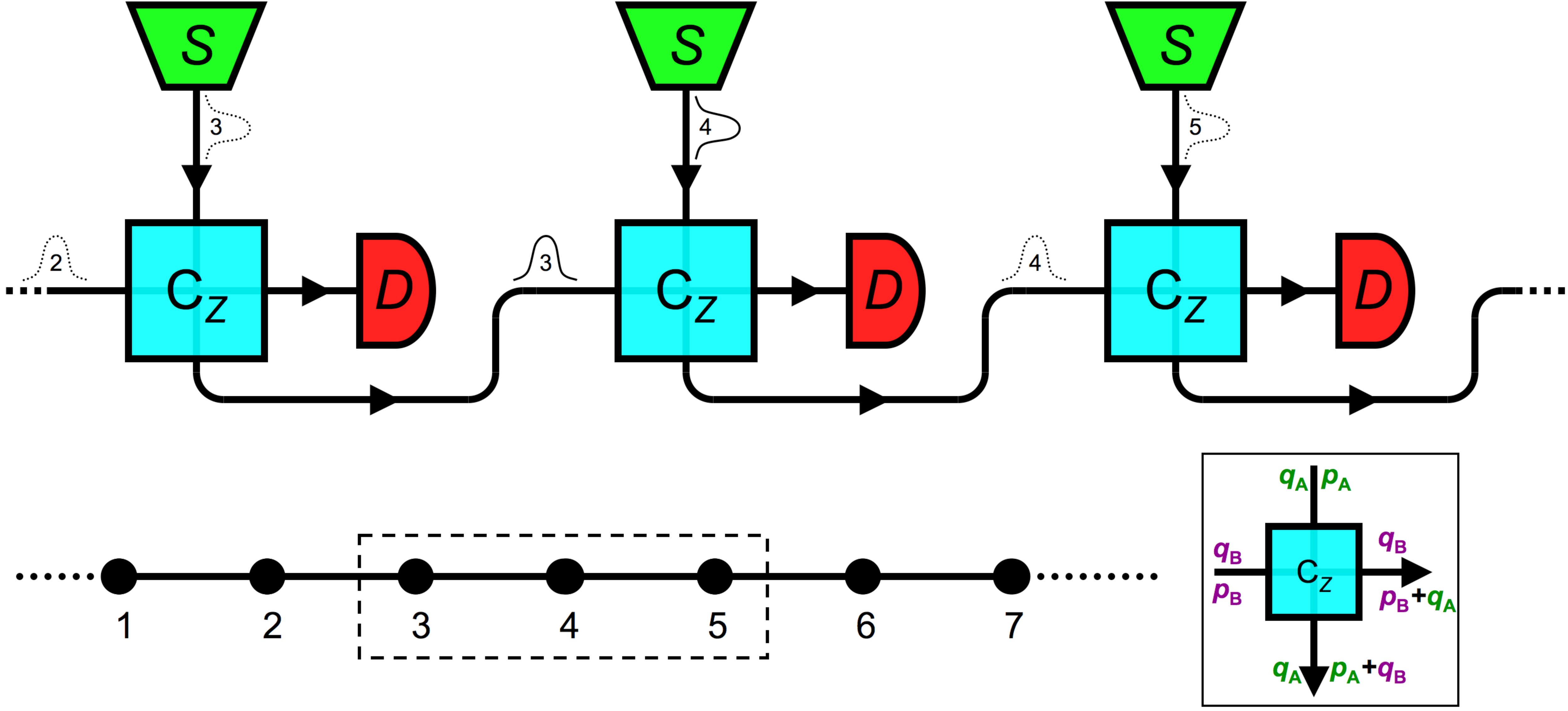}
\vskip -1em
\caption{Linear CV cluster state (quantum wire) made from squeezers and $\CZ$~gates.  The graph for a quantum wire of arbitrary extent is shown at the bottom, with the three highlighted nodes corresponding to the partial experimental schematic above.  Black lines represent optical paths (fiber or free space).  Single-mode squeezers~$S$ each produce a single pulse in a squeezed vacuum state (reduced variance in~$\hat p$), timed so that they intersect the previous and subsequent pulses within the appropriate $\CZ$~gates.  The Heisenberg action of $\CZ$ on the input quadratures is shown diagramatically in the box at the bottom right ($\hat p_A \to \hat p_A + \hat q_B$ and $\hat p_B \to \hat p_B + \hat q_A$, with~$\hat q_A$ and~$\hat q_B$ unchanged).  Small numbered pulses are drawn next to the optical paths to illustrate their positions at a particular instant during the experiment.  Dotted pulses show their locations (and their labels) at previous and subsequent times.  Notice that pulses~3 and~4 are each shown twice, illustrating their flow through the experiment.  Detectors~$D$ implement the desired quantum algorithm through adaptive measurements that depend both on the algorithm and on previous measurement results~\cite{Gu2009}.  We are ignoring boundary conditions (see text), so the quantum wire is effectively infinite, and the schematic shown is taken to be repeated indefinitely both to the left and right.}
\vskip -1  em
\label{fig:linearQNDextended}
\end{center}
\end{figure}

\begin{figure}
\begin{center}
\includegraphics[width=.7 \columnwidth]{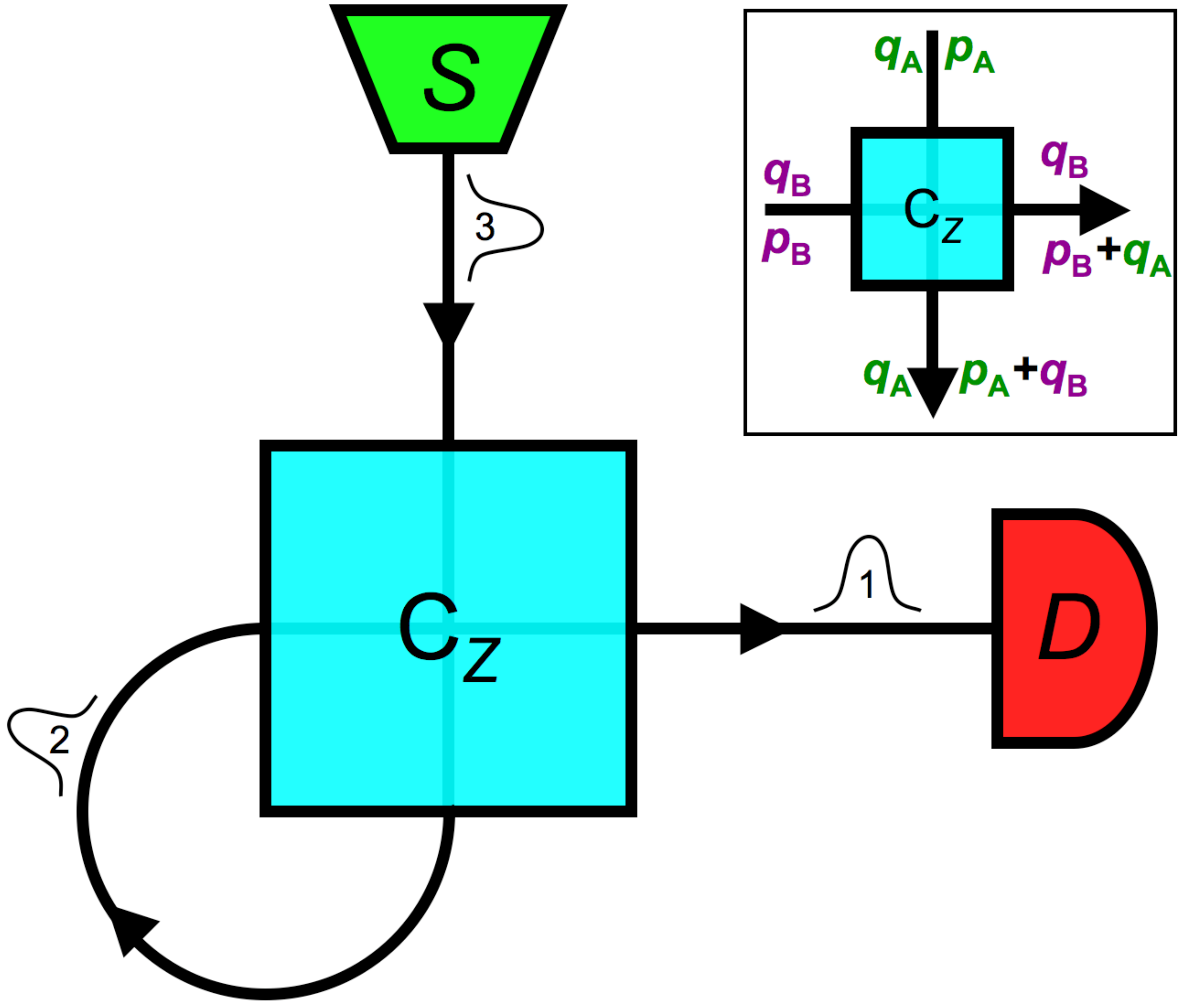}
\vskip -1em
\caption{Linear CV cluster state made from just one squeezer and one $\CZ$~gate.  The setup from Figure~\ref{fig:linearQNDextended} is ``rolled up'' into a compact design.  The squeezer~$S$ continually generates pulses of $\hat p$-squeezed vacuum at regular intervals, with the length of the looping path chosen so that adjacent pulses are modematched and phaselocked within the $\CZ$ gate.  The Heisenberg action of this gate is indicated in the box and is the same as in Figure~\ref{fig:linearQNDextended}.  The detector~$D$ measures the pulses in sequence as they exit the $\CZ$ gate, with current measurement results affecting future measurement bases~\cite{Gu2009}.  The numbered pulses next to the optical paths indicate the position of the independent modes in a single snapshot of the experiment, with pulse~1 having been emitted from~$S$ before pulse~2,~etc.  Because the mode pulses are distinguished by the time of their emission from~$S$, we say that this experiment generates a temporal-mode CV quantum wire.}
\vskip -1  em
\label{fig:linearQNDtemporal}
\end{center}
\end{figure}

\emph{Quantum wire.}---%
We illustrate the basics of our approach with a one-dimensional cluster state, which corresponds to a line graph and is also known as a \emph{quantum wire}.  Such a cluster state can be used to perform single-mode operations on a single qumode~\footnote{Short for ``quantum mode''---used in analogy with \emph{qubit} to represent one abstract unit of CV quantum information.} of quantum information~\cite{Gu2009}.  The canonical method of constructing CV cluster states involves linking single-mode squeezed states together using $\CZ$~gates~\cite{Menicucci2006,Gu2009}.  One possible setup for an optical experiment producing and using an $N$-node quantum wire is shown in Figure~\ref{fig:linearQNDextended}.   (We purposefully neglect boundary conditions at the ends of the cluster state for the time being---we will deal with them later.)  While this is not the easiest experimental proposal in light of the other methods available~\cite{vanLoock2007,Ukai2010}, we use it as a pedagogical tool to make the leap to a temporal-mode encoding and illustrate its power.

In any optical experiment using CV cluster states, the optical modes corresponding to each node must be specified.  There are many ways to do this since there are many ways to decompose the electromagnetic field into independent quantum modes.  Infinite plane waves---often used in quantum field theory---are distinguished on the basis of frequency, direction of propagation, and polarization, while experiments in quantum optics often use localized modes in the form of finite pulses~\cite{Bachor2004}, which are distinguished by their shape and location in spacetime.  In Figure~\ref{fig:linearQNDextended}, the output of the single-mode squeezers~$S$ are assumed to be similarly-shaped pulses of finite duration (and thus also of finite spatial extent).  They are localized compared to the distance between optical elements, and they are distinguished by their location in space at a given time.  The pulses from the squeezers must be modematched and phaselocked within each of the $\CZ$~gates.  The measurements performed by the detectors~$D$ are chosen adaptively based on the desired single-mode operation to be performed and previous measurement outcomes~\cite{Gu2009}.

Notice that the squeezers~$S$ emit pulses sequentially so that each of them reaches the $\CZ$~gate from the top simultaneously with the previous one, which enters from the left.  The diagram is rather repetitive, though, and one might wonder whether it would be possible to reuse some of the optical elements.  Specifically, one might be tempted to connect the dotted end of the path on the right to the one on the left and just have each squeezer fire multiple times.  In fact, if this were done, no more than one copy of each optical element would be needed, with the individual modes now all occupying the same optical paths but at different times---a \emph{temporal} encoding of mode pulses.  Our proposal is exactly this, and the resulting setup is shown in Figure~\ref{fig:linearQNDtemporal}.

Notice that we need not care that the first node in the experiment gets coupled to vacuum (instead of to squeezed vacuum) by the $\CZ$~gate---we can simply measure $\hat q$ on the first node (which interacted with a vacuum-state ``zeroth node'' within the $\CZ$~gate the first time around) to delete it from the graph~\cite{Gu2009}.  The computation can then proceed normally, starting with the second node.  Also notice that we do not ``write'' quantum information into the beginning of the quantum wire, starting instead with an ``empty'' quantum wire~\cite{Gu2009}.


\begin{figure}
\begin{center}
\includegraphics[width=\columnwidth]{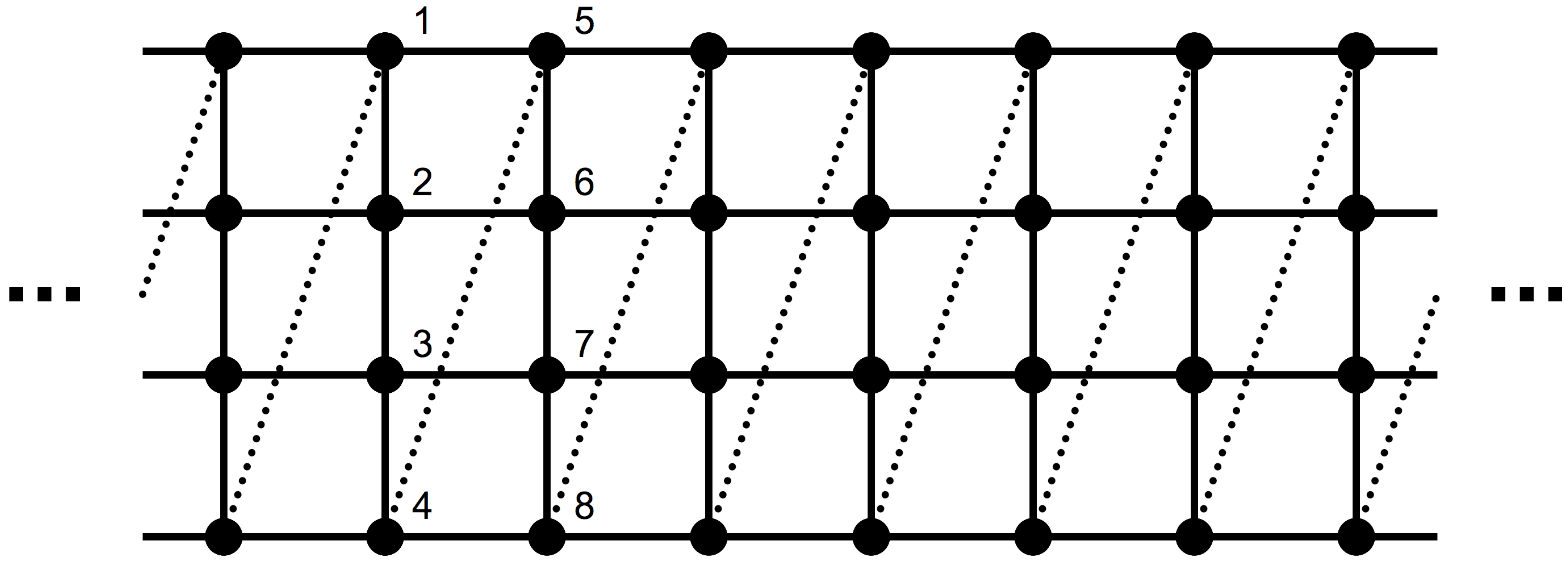} \\[1.5em]
\includegraphics[width=\columnwidth]{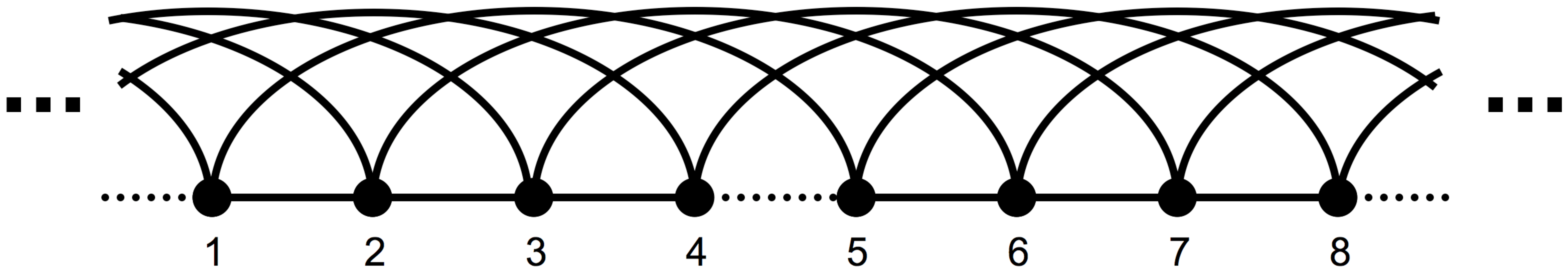}
\vskip -1em
\caption{Two-dimensional square-lattice graph (top) that is infinite in one dimension but finite in the other.  This graph can be redrawn as a multiply threaded infinite line graph (bottom).  There are $M$~additional threadings of the line that pass through nodes $M$~units apart, resulting in a square lattice with vertical dimension~$M$.  (In this case, $M=4$.)  The dotted links represent additional $\CZ$~interactions that would make the linear version translationally symmetric and equivalent to a square lattice on a cylinder with one unit of shear in the longitudinal direction.  Such a family of graphs would still be universal for one-way QC because we can measure~$\hat q$ on every $M$th~node to delete it (and its links) and ``unfold'' the graph into an ordinary square lattice with a vertical dimension of \mbox{$(M-1)$}~\cite{Gu2009}.}
\vskip -1  em
\label{fig:2dQNDlinearized}
\end{center}
\end{figure}

\begin{figure}
\begin{center}
\includegraphics[width=\columnwidth]{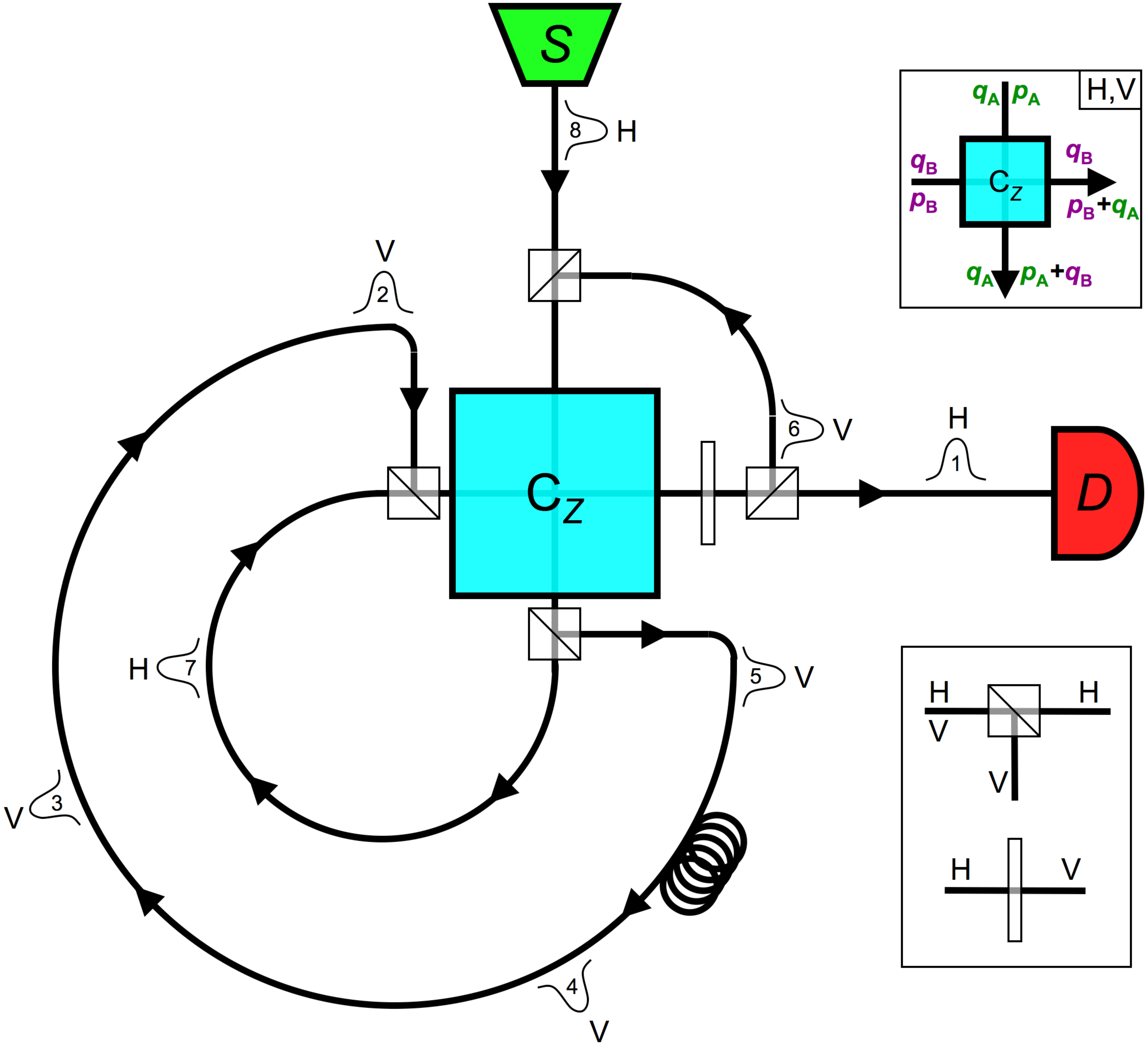}
\vskip -1em
\caption{Two-dimensional square-lattice CV cluster state made from just one squeezer and one $\CZ$~gate.  Beginning with the setup in Figure~\ref{fig:linearQNDtemporal}, which is assumed to use horizontal polarization exclusively, polarizing beamsplitters and a half-wave plate (their action shown in the box at the bottom right) divert the output of the first pass through the $\CZ$~gate (inner loop) around for a second pass (outer loop) before heading to the detector~$D$.  This provides the additional ``threadings'' required for a square lattice (Figure~\ref{fig:2dQNDlinearized}, dotted links included).  The $\CZ$~gate must act in the same fashion when the input states are horizontally polarized as when they are vertically polarized (see box at the top right)~\cite{endnote30}.  The coil in the larger loop indicates that the length of this loop controls the vertical width of the lattice (the parameter~$M$ from Figure~\ref{fig:2dQNDlinearized}), with $M=4$~shown.  The numbered pulses next to the optical paths indicate the position and polarization of the independent modes in a single snapshot of the experiment, with pulse~1 having been emitted from~$S$ before pulse~2,~etc.  Because the mode pulses are distinguished by the time of their emission from~$S$, we say that this experiment generates a temporal-mode CV cluster state.}
\vskip -1  em
\label{fig:2dQNDpolarization}
\end{center}
\end{figure}

\emph{Square lattice.}---%
Line graphs are useful for manipulation of a single qumode of quantum information.  For universal one-way QC using CVs, however, such a graph is insufficient~\cite{Gu2009,Nielsen2006}, and a two-dimensional graph of sufficient connectivity, such as a square lattice, is required.  In most cases, including the original proposal for qubit-based one-way QC~\cite{Raussendorf2001}, such a graph is thought of as consisting of many quantum wires arranged vertically, with the computation proceeding from left to right.  Vertical links between wires allow for two-mode operations ($\CZ$~gates).  Normally, a fully connected square lattice is neither necessary nor desired for a given computation.  Extra nodes and their links may be deleted by measuring them in the $\hat q$-basis~\cite{Gu2009}, allowing for more useful graphs to be lifted from the universal square lattice.

Because the computation is usually pictured to proceed from left to right with bounded vertical width, we consider a square lattice graph that is formally infinite in the horizontal dimension but finite in the vertical.  (Once again, we ignore the boundary issues associated with a finite graph for now.)  Two forms for this graph are shown in Figure~\ref{fig:2dQNDlinearized}.  The linearized form is useful for our purposes because we can start with the experimental setup in Figure~\ref{fig:linearQNDtemporal}, which generates an infinite quantum wire, if we can modify it to include the additional threadings required.  This can be achieved by feeding the outgoing modes from the setup in Figure~\ref{fig:linearQNDtemporal} back into a second $\CZ$~gate before letting them enter the detector.  This $\CZ$~gate should be identical to the first but with a longer looping path, allowing nonadjacent nodes to be linked together.  The train of output modes would then be sent to the detector.  In fact, we can go one step further and reuse the same $\CZ$~gate if we can multiplex both pulse trains in the same optical path.

One way to do this is shown in Figure~\ref{fig:2dQNDpolarization} and requires that the $\CZ$~gate act in the same fashion on horizontally polarized inputs as on inputs polarized vertically~\footnote{Whether this is technically two $\CZ$~gates instead of one is a matter of implementation.  This factor of two does not concern us---both are a drastic  improvement over~$O(N^2)$.}.  Alternatively, we could use actively controlled beam diverters in place of the polarizing beamsplitters to multiplex the second pass of the modes through the $\CZ$~gate as long as we ensure the two pulse trains never overlap.  This would allow us to use a $\CZ$~gate that works with only a single polarization, with the cost being active control of the beam diverters instead of (passive) polarizing beamsplitters.  In either case, the output is a train of pulses, all with equal polarization, arranged in the order shown in Figure~\ref{fig:2dQNDlinearized} (dotted links included).  The pulses head to the detector in sequence, striping vertically down each line in sequence from left to right.  Even though the graph has infinite horizontal extent, at no time are there more than a fixed number of unmeasured modes present.  In this way, more cluster state is continually generated as previous modes are consumed by the detector.

The solution for vacuum-state contamination of the beginning of the lattice is analogous to that for the quantum wire: $\hat q$\mbox{-}measurements on the first $M$~nodes (i.e.,\ the first vertical stripe) delete these nodes and their links from the graph, thereby isolating the usable CV cluster state from the extra vacuum-state nodes linked in at the beginning.  Once again, we do not ``write'' quantum information into the beginning of the cluster state, starting instead with an ``empty'' one-way quantum computer~\cite{Gu2009}.

These cluster states can be used for universal one-way QC using CVs.  For general Gaussian operations, including multimode Gaussians, homodyne detection alone is sufficient~\cite{Gu2009}, with the basis chosen by phase-shifting a local oscillator~\cite{Bachor2004}.  If universal QC is desired, then we only need the ability to perform photon counting, as well~\cite{Gu2009}.  Thus, there are only two detector setups required, along with a controllable beam diverter to select either the homodyne detector or photon counter.  In addition, the pulses must be separated by enough time to do the classical feedforward of previous measurement results and prepare the detector for the next measurement before the next pulse arrives~\cite{Gu2009}.  Because only a finite portion of the cluster state exists at any given instant, the requirements for coherence and stability do not increase with the length of the computation, although they will increase with the vertical width of the cluster state.  Finally, it has been known since the beginning~\cite{Menicucci2006,Gu2009} that CV cluster states are fundamentally imperfect due to the fact that finite energy implies finite squeezing and that this results in errors in the computation.  Recent results~\cite{Ohliger2010} have reiterated the need for a comprehensive approach to fault tolerance within the CV one-way QC platform.  This issue affects all proposed implementations and remains the subject of ongoing research.


\emph{Conclusion.}---%
The canonical method for generating continuous-variable cluster states was quickly dismissed as too resource intensive and difficult because of the experimental challenges associated with quantum nondemolition (QND)~gates, of which the controlled-$Z$ ($\CZ$)~gate is one example.  These gates have one advantage over most other entangling gates, however: they all commute with each other.  This places the $\CZ$~gate in a privileged position with respect to extensible experimental design.  We use this feature to design an experiment that can be used for an arbitrarily long one-way quantum computation but that requires only one $\CZ$~gate.  This dramatically changes the resource requirement landscape for continuous-variable cluster states, now allowing experimentalists to focus on perfecting just a \emph{single copy} of the necessary optical elements---single-mode vacuum squeezer, $\CZ$~gate, homodyne detector, and photon counter---since we have shown here that one copy of each is all that is needed for universal one-way quantum computation.


We thank Peter van~Loock and Steve Flammia for discussions.  N.C.M.~is grateful for support from The University of Queensland and the Centre of Excellence for Quantum Computer Technology for a visit that made this work possible.  Research at Perimeter Institute is supported by the Government of Canada through Industry Canada and by the Province of Ontario through the Ministry of Research \& Innovation.

\vskip -1em


\bibliographystyle{bibstyle_notitle}
\bibliography{TemporalCVCS}

\end{document}